\documentclass[twocolumn,aps,showpacs,floatfix,amssymb,preprintnumbers,superscriptaddress,showkeys,longbibliography]{revtex4-2}
\usepackage[latin9]{inputenc}
\usepackage{amsmath}
\usepackage{graphicx}

\makeatletter
\usepackage{dcolumn}
\usepackage{bm}
\usepackage{epstopdf}
\usepackage{url}
\usepackage{xcolor}
\usepackage{xcolor}

\makeatother

\begin{document}
\title{Density fluctuations in granular piles traversing the glass transition: A grain-scale characterization of the transition via the internal energy.}
\author{Paula A. Gago}
\email{paulaalejandrayo@gmail.com}
\affiliation{Department of Earth Science and Engineering, Imperial College, London,
SW7 2BP, UK.}
\author{Stefan Boettcher}
\affiliation{Department of Physics, Emory University, Atlanta, GA 30322, USA}
\email{sboettc@emory.edu}

\begin{abstract}
The transition into a glassy state of the ensemble of static, mechanically stable configurations of a tapped granular pile is explored using extensive molecular dynamics simulations. We show that different horizontal sub-regions ("layers") along the height of the pile traverse this transition in a similar manner but at distinct tap intensities. We supplement the conventional approach based purely on properties of the static configurations with investigations of the grain-scale dynamics by which the tap energy is transmitted throughout the pile. We find that the effective energy that particles dissipate is a function of each particle's location in the pile and, moreover, that its value plays a distinctive role in the transformation between configurations. This internal energy provides a "temperature-like" parameter that allows us to align the transition into the glassy state for all layers, as well as different annealing schedules, at a critical value.
\end{abstract}

\maketitle

\begin{figure*}
\centering \includegraphics[width=1\textwidth]{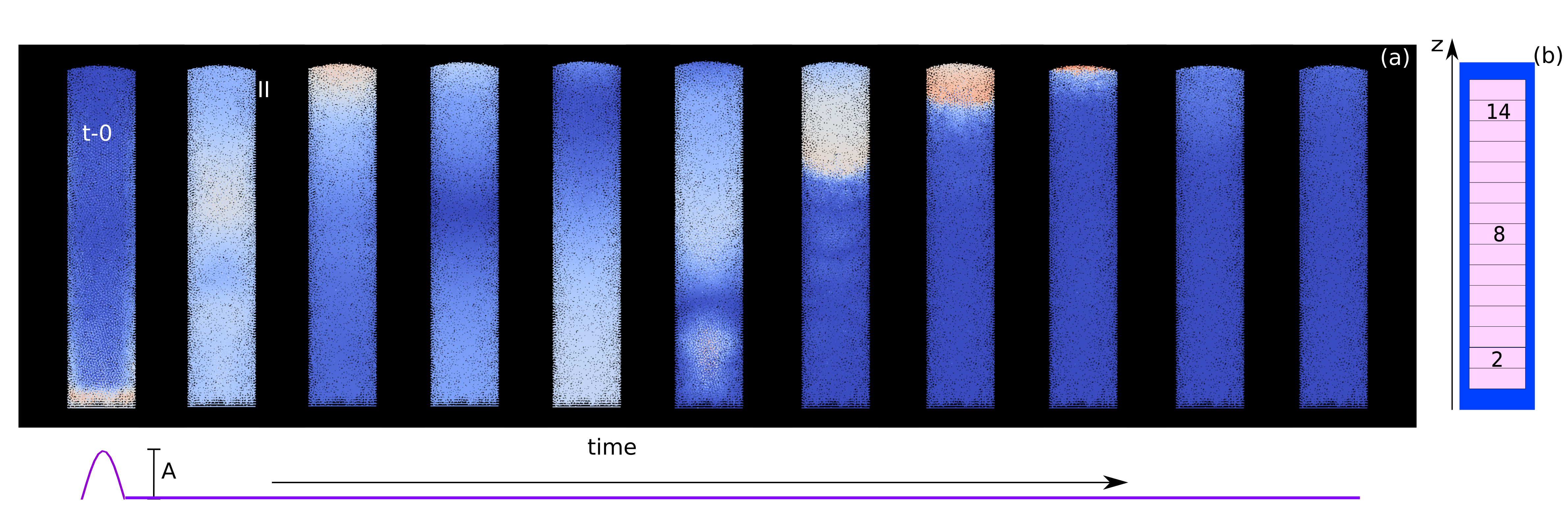} 
\vskip -0.3cm
\caption{{\bf Evolution of a single tap.} (a) Snapshots of the cylindrical silo illustrating the pile dynamics during a single tap. 
At time $t=0$, the tap is applied (left panel). The purple line below the snapshots shows the schematic of the corresponding perturbation.
The kinetic energy injected traverses the pile (panels from left to right) until it is fully dissipated through collisions and friction. The color 
scale reflects the instantaneous speed of each particle, increasing from blue over white to red. (b) A schematic of the $15$ disk-like 
sub-regions ("layers") used to measure the local packing fraction, with three representative layers (for low, middle and top) highlighted for 
future reference. }
\label{fig:setup}
\end{figure*}

\section*{Introduction}\label{Introduction}
Granular materials are ubiquitous in nature and have always fascinated scientists, Coulomb and Reynolds among them, due to their
somewhat counter-intuitive behavior \cite{de1999granular}.  
They have been  considered as an additional state of matter in its own right \cite{jaeger1996granular} and although they are the second-most common form by which mankind handles
materials (behind fluids), the way in which mechanical perturbations define its macroscopic properties are still not well understood. 
One of their key characteristic is that they tend to be at rest: Any input of energy provided to the system , whether by tapping, shearing, or tilting, is eventually 
dissipated through frictional contacts and collisions among the grains. Specifically, it requires such an external input of energy to be able to 
modify its configuration. Thus, the way in which the system behaves (fluid, solid, gas) is strongly determined by the way this external energy 
is dispensed and the microscopic processes leading to its dissipation. Attributes such as friction, density, the granular contact network, 
container geometry, etc, will likely play a role in this process. 

Although their propensity to stay at rest marks these materials as ``athermal'', any boundary making it distinct from other, thermal states of matter are progressively weakening. For example, we have recently described the transition of a tapped granular pile into a glassy state \cite{gago2020universal} familiar from polymers, complex fluids, and frustrated magnets \cite{Hutchinson1995,Debenedetti01,Zhu13,roth2016polymer,Gedde19, Sibani06a}. For such disordered thermal materials, relaxation times increase for many orders of magnitude (and possibly even diverge) over a small range of temperatures \cite{Debenedetti01}, beyond which the systems remains out of equilibrium for any practical purpose. Whether the transition was approached gradually or via a hard quench in the intensity of taps, we encountered  the same phenomenology in the granular pile as is observed for other glass-forming materials.  Although the distribution of fluctuations elicited by a perturbation within a granular medium may have profound differences to conventional thermal noise \cite{nowak1998density}, as long as sizable fluctuations exist to activate events, glassy relaxation appears to be universal \cite{robe2016record,boettcher2021extreme}.

In 1989 Edwards proposed the possibility of employing the formalism of statistical mechanics
to describe the properties of static granular materials \cite{edwards1989theory}. His theory considered as the statistical ensemble the set of 
static, mechanically stable configurations, that the system acquires after having dissipated the kinetic energy received by repeated 
perturbation. It set a milestone in the study of granular materials, as it hinted at some order behind the disorder. Since then, many studies  
\cite{nowak1998density,blumenfeld2009granular,makse2002testing,puckett2013equilibrating,schroter2005stationary,pugnaloni2011master,pugnaloni2010towards,gago2016ergodic,henkes2007entropy,bililign2019protocol} have addressed these questions.

One of these studies is the well known Chicago experiment \cite{nowak1997reversibility}, which presented a simple perturbation protocol 
able to create such a collection of states. This protocol consists of the repeated application of discrete ``taps'' to a granular pile confined 
inside a container and the collection of static configurations obtained after the system has dissipated the injected energy. 
For taps with low (fixed) intensities, the system shows a logarithmic increase in density, or packing fraction $\phi$, with the number of applied taps \cite{nowak1998density,gago2020universal, sibani2016record, richard2005slow}. However, executing a 
series of stepped annealing protocols, both for increasing and decreasing intensities of the perturbation, the system undergoes a fast 
``irreversible'' transient of low densities and reaches a so-called ``reversible regime''. There, the density of the static configurations becomes a function only of 
the intensity of the  tap applied. The collection of static configurations at a given tap intensity serves as a setting to test Edwards' hypothesis.

Employing a protocol similar to stepwise annealing used in Ref.~\cite{nowak1997reversibility}, but decreasing the tap intensity (represented conventionally by its reduced acceleration $\Gamma$) 
continually between taps at various variation rates ($\dot\Gamma$), we have shown \cite{gago2020universal} that the packing fraction $\phi$ as a function of 
$\Gamma$ behaves in a manner resembling the glass transition found in thermal materials    
\cite{gedde2019fundamental,hunter2012physics,fischer_hertz_1991,Debenedetti01}. That is, at high $\Gamma$, $\phi(\Gamma)$ evolves independent of 
$\dot\Gamma$, while for lower intensities it splits into a separate branch for each $\dot\Gamma$, reaching higher densities for lower $\dot\Gamma$.

Here, we explore the origin of that 
transition in greater detail, accounting for the heterogeneous response of the system at different heights of the pile to the same tap \cite{nowak1998density, gago2015relevance, mehta2008heterogeneities}, as imposed by gravity.
As different layers attain different densities along the same protocol, and enter a glassy state at different tap intensities 
$\Gamma$, we find that a grain-level examination of the energy propagation along the system \emph{during} a tap allows us to collapse the density behavior for the different layers 
when plotted as a function of the effective energy grains receive from the collective perturbation. In particular, we find a critical value of this 
effective energy along which the peaks exhibited by the density fluctuations in each layer align. 

In effect, we are taking back a step from the macroscopic perspective of Edwards' hypothesis, especially its focus on volume assuming the 
role of the controlling parameter \cite{edwards1989theory} (comparable with temperature in equilibrium statistical mechanics). Instead, we delve into a microscopic analysis of the 
dynamic process that leads from one static configuration to the next. 
It opens the door to a first-principles, grain-level characterization of the 
impact a perturbation has on granular systems, in the spirit of a kinetic theory in statistical physics relating temperature to internal energy. 
Hence, this microscopic perspective holds the promise to be generalizable to other perturbation protocols (such as seismic or acoustic waves, avalanches, etc.) as well as to systems with different geometries and grain properties. 
Although our investigations as-of-yet fall short in understanding the full impact this effective energy has on transforming configurations, 
aligning data according to it already explains, e.g., unusual behavior of critical density fluctuations observed previously for the pile 
as a whole \cite{schroter2005stationary}. 

For these insights, we had to perform detailed and extensive molecular dynamics simulation (MD) of soft-
spheres, in particular using the implementation provided by $LIGGGHTS$ \cite{goniva2012influence} open source software, to record internal energies during the dynamic process as well as to sample with sufficient statistics for macroscopic variables of static configurations, such as the density and its fluctuations in each layer of the pile.

\begin{figure*}
\centering \includegraphics[width=1\textwidth]{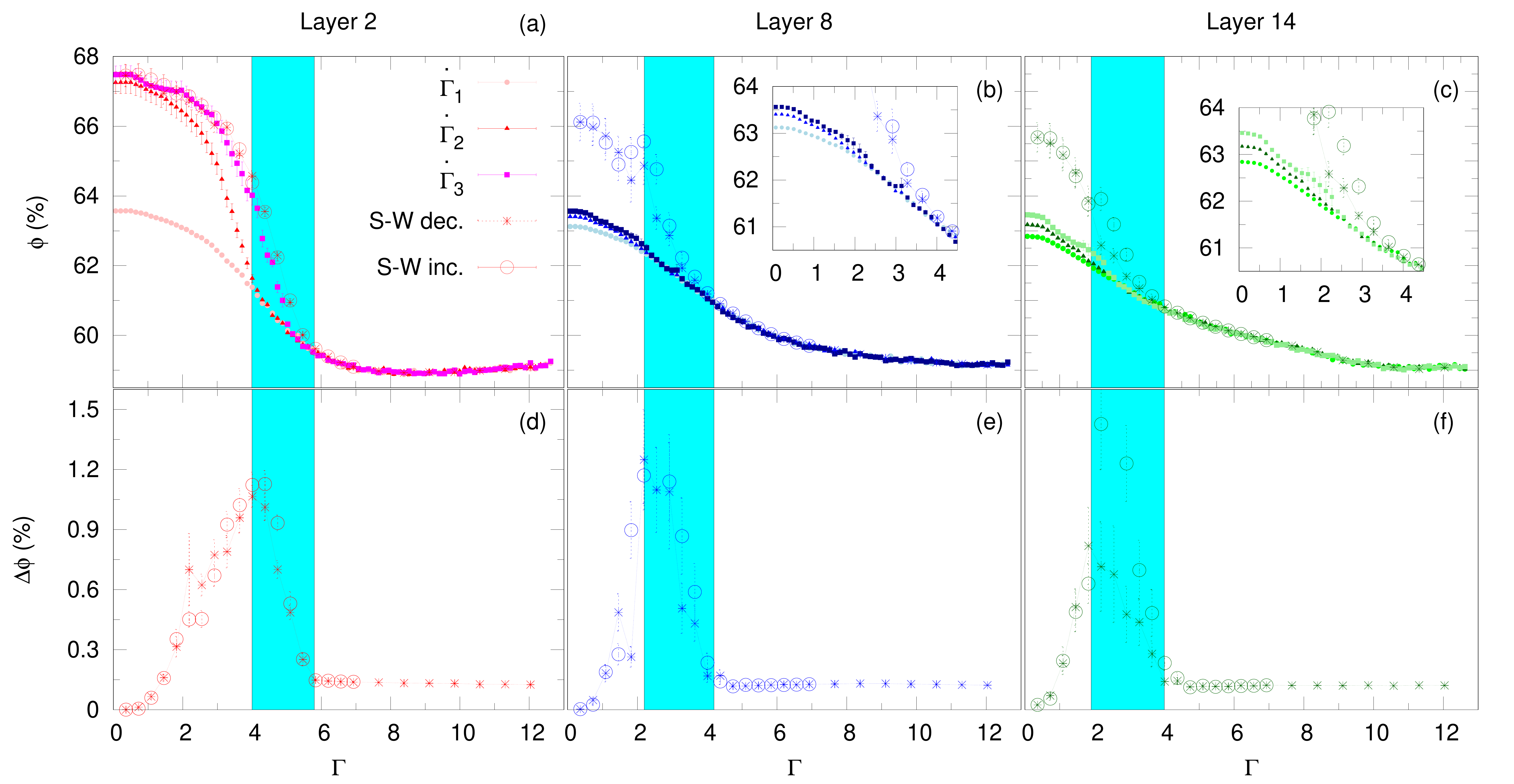} 
\caption{{\bf Packing fraction and density fluctuations in the pile.} (a)-(c) Packing fraction $\phi$ as a function of $\Gamma$ for layers $2$, $8$ and $14$, respectively, located at different heights of the pile, 
marked in Fig.~\ref{fig:setup}(b). For the continuous protocol it is $\dot\Gamma_{1}=0.00002{\rm m}(\omega^{2}/g)$ per tap (red circle), while
$\dot\Gamma_{2}=\dot\Gamma_{1}/4$ (blue triangle-up) and $\dot\Gamma_{3}=\dot\Gamma_{1}/8$ (green squares). For the 
stepwise protocol, crosses correspond to the decreasing protocol and open circles to its reverse, i.e., increasing protocol.  Insets merely enlarge main plot for $\Gamma<4$.
(d)-(f) Density fluctuations $\Delta \phi$ as a function of $\Gamma$ for the same three layers displayed in (a)-(c), respectively, as obtained 
by the step-wise protocol. For each protocol, $16$ independent realizations were performed, and error bars in all the figures correspond to 
the standard error of their mean.
Shaded areas mark the $\Gamma$-range corresponding to the right slope of each peak in $\Delta \phi$ vs. $\Gamma$.}
\label{fig:layers_gamma}
\end{figure*}

Figure~\ref{fig:setup} serves to illustrate the setup for the granular pile used in our MD simulations. 
Specifically, Fig.~\ref{fig:setup}(a) presents a series of snapshots of the pile describing the dynamic process during a single tap, after the perturbation has been exerted at $t=0$ (left-most panel). The following panels of Fig. \ref{fig:setup}(a) show the kinetic energy (color coded from blue for $v=0$ 
over white to red for high speeds $v$) provided by the tap ``spreading''
through the pile and finally getting fully dissipated.
The dimensionless acceleration $\Gamma$ is used to represent the tap intensity.
 
As can be seen from Fig.~\ref{fig:setup}(a), the energy imparted by the tap travels along the height of the pile neither in an instantaneous 
nor homogeneous fashion. Instead, a complex process of energy transfer and dissipation ensues under the influence of gravity that we 
intend to study in more detail below. It is therefore not surprising that the granular density of static configurations is often measured over  
narrow layers of constant height \cite{nowak1997reversibility,gago2015relevance, mehta2008heterogeneities}. 
To measure the packing fraction $\phi$, we divide the \emph{entire} system into $15$ cylindrical sub-regions ("layers"), as schematized in Fig.~\ref{fig:setup}(b), stacked along the height of the pile.

In our simulations, we start from a high tap intensity and implement two distinct annealing 
protocols of decreasing intensity, one continuous and the other stepwise. For the first protocol, three different continuous rates of change
were used.
The packing fraction $\phi$ of each layer is measured after each tap.
In order to measure also the density fluctuations $\Delta\phi$, corresponding to the ``stationary state'' at a given tap intensity, a stepped protocol 
inspired by Ref.~\cite{nowak1997reversibility} is performed. In this protocol we apply a series of taps at each intensity $\Gamma$ 
before decreasing it by decrementing the tap amplitude by an amount $\Delta A$.  
Finally, to verify the stationarity and reversivility of the produced states, the stepwise protocol was repeated in the reversed direction, i.e., for increasing intensities.

\section*{Discussion of Results}\label{Discussion}
\paragraph*{Local density and its fluctuations:}
Figure~\ref{fig:layers_gamma}(a)-(c) shows $\phi$ as a function of $\Gamma$ for three $\dot\Gamma$ under continuous annealing 
for each one of the three layers located at the low, middle and top of the pile marked in Fig.~\ref{fig:setup}(b). As previously reported \cite{gago2020universal}, 
within each fixed layer, it can be seen that the data obtained at higher intensities $\Gamma$ vary together, irrespective of $\dot\Gamma$, 
consistent with equilibrium behavior. However, for lower intensities,  they split off  into separate branches for different $\dot\Gamma$. As a 
consequence, the final density achieved for $\Gamma\to0$ becomes a function of the protocol. Significantly, the regime of $\Gamma$ 
where this split occurs itself depends on the layer, suggesting that each strata in the pile transitions into a glassy state at a different 
tap intensity. Variations in the state between different parts of the pile due to gravity under the same perturbation has been noted previously 
\cite{mehta2008heterogeneities}. As we will show below, these variations can be accounted for by the difference in the effective energy acting in each region.

Also shown in Fig.~\ref{fig:layers_gamma}(a)-(c) is the density as a function of $\Gamma$ obtained in the stepwise protocol. As many taps 
are spent at fixed values of $\Gamma$ (and many more overall compared to any of the continuous protocols), the densities for this protocol 
are well-converged and provide an upper bound on the faster-moving, continuous protocols. Starting at high $\Gamma$, crosses mark the decreasing protocol while open circles correspond to the same protocol but reversed after the 
decreasing protocol is completed at $\Gamma \approx 0.37$. Both set of data demonstrate that there is only a minute aging effect on its reversibility 
that can be ignored for our purposes here. 

Focusing on the corresponding density fluctuations  $\Delta \phi$ obtained in the stepwise protocol, as shown in 
Fig.~\ref{fig:layers_gamma} (d)-(f), we find that irrespective of the direction of the protocol, the fluctuations exhibit identical features as a function of the tap intensity (with small quantitative differences in strength). It 
can be observed that $\Delta \phi$ as a function of $\Gamma$ remains almost constant for high intensities until it sharply peaks for lower 
$\Gamma$ values, before vanishing as the intensity further decreases. This single-peaked behavior for each layer is consistent with 
previously reported results \cite{pugnaloni2011master, gago2015relevance}. However, we note that the right slope of each peak can be associated with the density transitioning into a glassy state, as it corresponds to the regime of intensities (highlighted by a blue stripe) where the splitting takes place in 
Fig.~\ref{fig:layers_gamma}(a)-(c) for the respective layer. (This behavior is independent of the direction of the protocol, showing that it is a 
function of the tap intensity and not a results of a residual transient.) Thus, although the rise in fluctuations and the onset of glassy behavior 
are aligned for each layer, neither the intensity $\Gamma$ of the macroscopic perturbation   
nor the respective densities, as would be expected from Edwards' hypothesis, allow to align the data for all layers simultaneously.  

\begin{figure}
\centering \includegraphics[width=0.9\columnwidth]{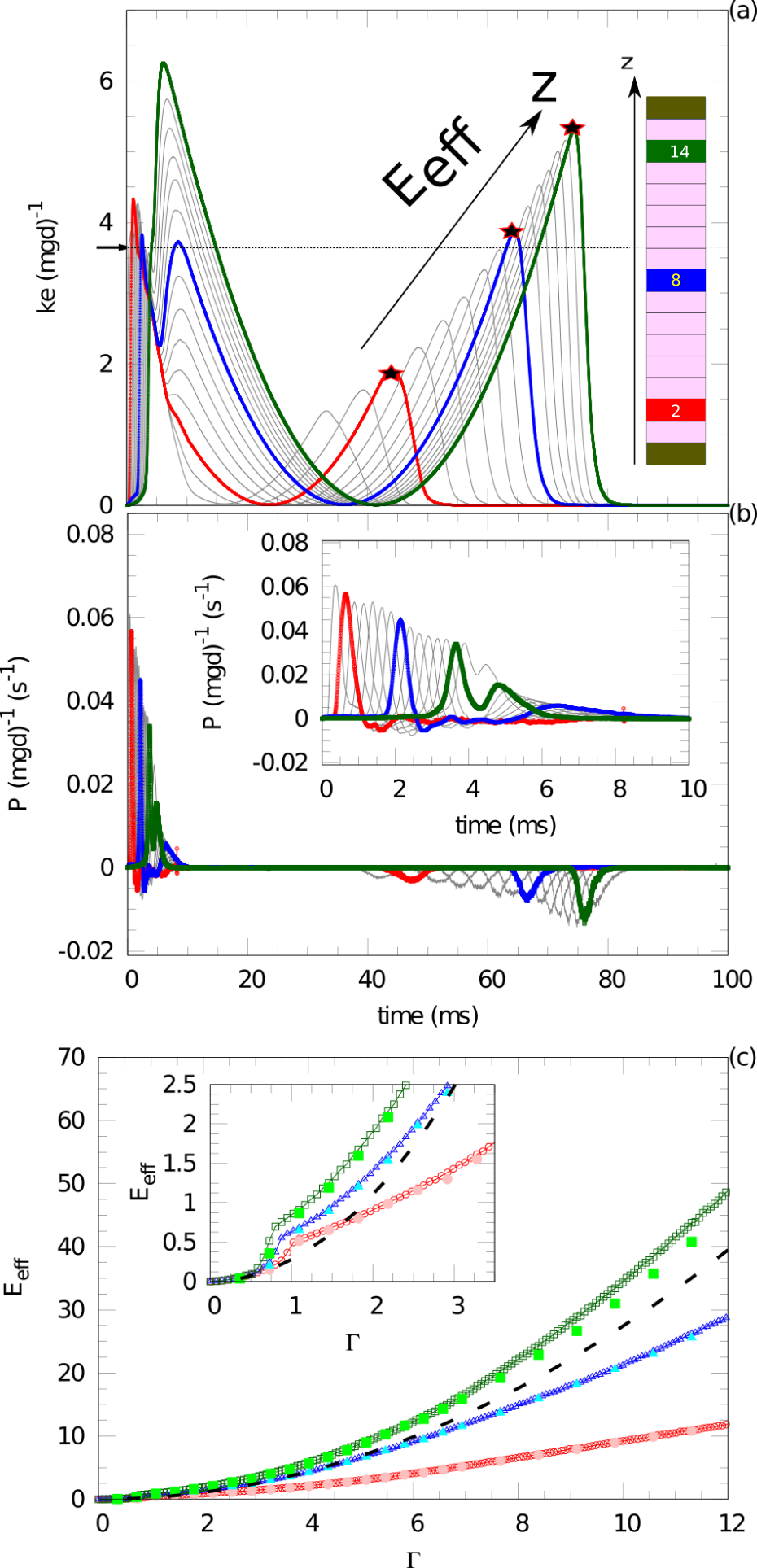} 
\caption{{\bf Energy transfer and dissipation during a tap.} (a) Average kinetic energy $\left\langle ke\right\rangle $ (in units of $mgd$) for particles during the dynamic process (corresponding to a perturbation of amplitude $A=0.002$) as a function 
of time (in color for layers $2$ (red), $8$ (blue) and $14$ (green), as indicated by the stylized height chart in the inset, all other layers indicated in gray). 
Initially, for $t=0$, the tap (almost instantaneously but slightly delayed through the compression of layers below) 
accelerates particles in each layer upward to the first peak in the kinetic energy, consistent with $(A\omega)^{2}/(2gd)$ (marked by arrow).  
Black stars record the effective energy $E_{\rm eff}$ of the last peak which is dissipated by crashing into the layer below. (b) Same as in (a) but plotting the corresponding power $P$ in the transfer, gain or loss, of the total mechanical energy for each layer. 
(c) Relation between $E_{\rm eff}$ and the tap intensity $\Gamma$, as experienced by each layer for continuous and stepwise protocols 
(open and full symbols, respectively). The dashed line marks the relation for the entire pile to behave as a homogeneous solid. The inset 
shows the enlargement of the low-energy regime most relevant to the glassy 
behavior, where departure from plain solid behavior is most evident. }
\label{fig:kinetic_e} 
\end{figure}

\begin{figure*}
\centering \includegraphics[width=1\textwidth]{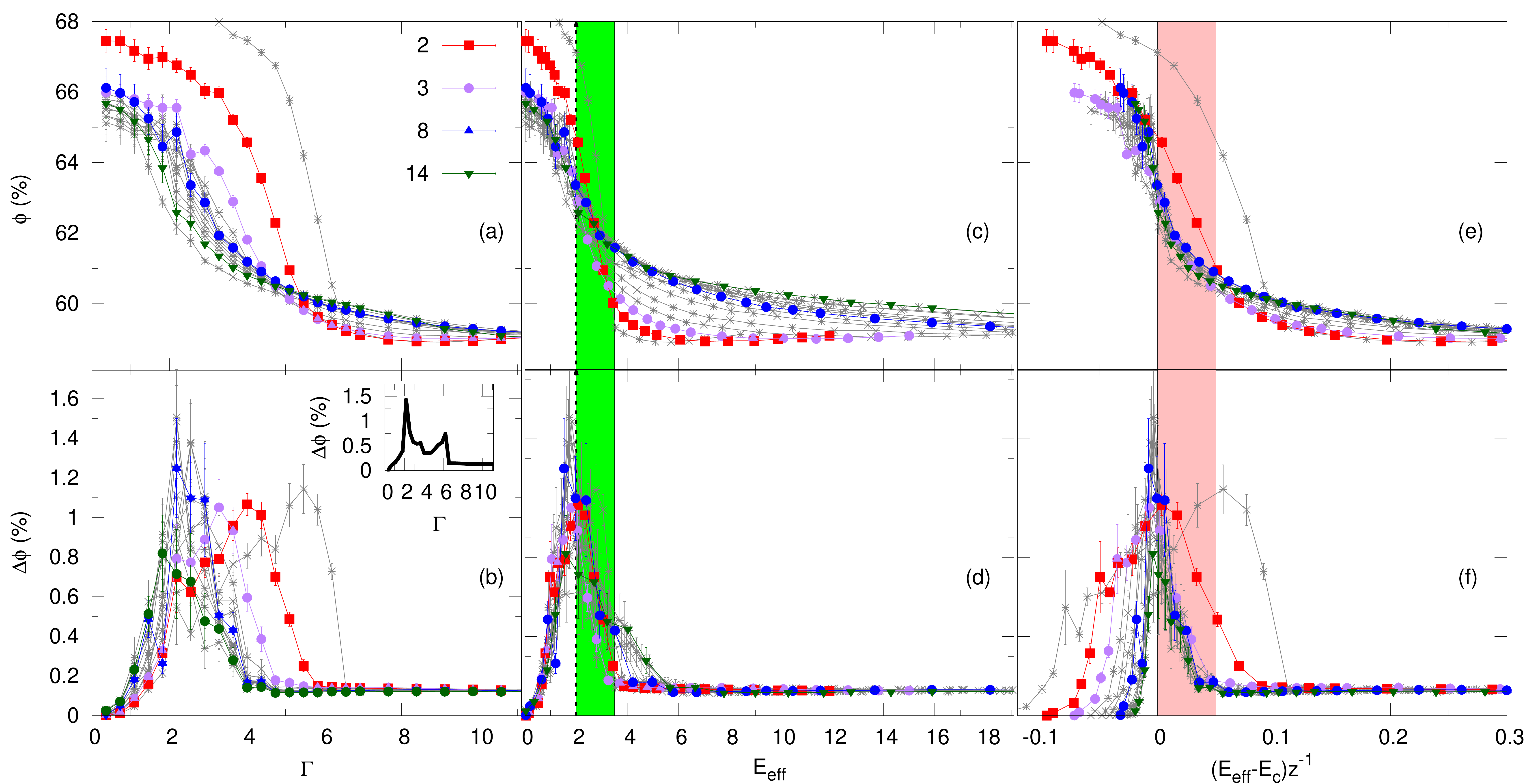}
\vskip -0.3cm
\caption{{\bf Data collapse for the stepwise protocol.} Plot of the density $\phi$ (a,c,e) and density fluctuations $\Delta\phi$
(b,d,f) obtained with the stepwise protocol. In panels (a-b), data is plotted as a function of $\Gamma$ and in panels (c-d)
as a function of $E_{\rm eff}$, measured for different layers along the
height of the granular pile. As a function of $\Gamma$, different layers behave similarly
but are shifted variably, without apparent collapse. In fact, when data
for $\Delta\phi$ is averaged over the layers, the individual peaks
convolve into a trough, see inset of (b). However, as highlighted in green in panels (c-d), the steep rise in density as well as the density 
fluctuations for all layers align when plotted as a function of $E_{\rm eff}$, revealing the signature of a glass transition (see also Fig.~\ref{fig:glass_transition}). Additional rescaling with the height $z$ of the pile improves the collapse in the density in panel (e) but also stretches the width of the fluctuations at low $z$. }
\label{fig:density} 
\end{figure*}

\paragraph{Effective energy of grains:}
To align the onset of glassy behavior between different layers, we now examine in more detail the dynamic process by which each tap 
distributes energy throughout the system. It permits us to identify an effective energy $E_{\rm eff}$ that quantitatively collapses the 
transition to glassy behavior for the different layers of the system. In particular, we find a specific threshold value $E_{\rm c}$ for which 
particles with $E_{\rm eff}<E_{\rm c}$  always assume glassy arrangements, apparently. To this end, we now look at the grain-scale 
dynamics of the energy transmission and dissipation during the tap.
 
Figure~\ref{fig:kinetic_e}(a) shows the kinetic energy $ke$ per particle (in units of $mgd$) as a function of time during the dynamics. In this example, 
we used an intermediate tap intensity of $\Gamma\approx3.65$ and averaged over particles belonging to the three representative layers 
along the pile. Each exhibits an initial peak, whose value is consistent with $(A\omega)^{2}/(2gd)\approx 3.65$, the kinetic energy of a 
particle moving with the maximum speed reached by the perturbation. After this initial peak and a brief period of energy transfer between 
layers, each layers follows a ``free fall'' behavior, decreasing its velocity until near zero before starting accelerating downwards and 
reaching a last peak. It is, in fact, the value of the last peak in $Ke$ that interests us, as this is the kinetic energy that particles 
dissipate when they settle. Marked by a black star, we will call this energy the effective energy $E_{\rm eff}$ that the perturbation imparts to 
each specific layer. It is easy to see that $E_{\rm eff}$ is an increasing function of the layer height $z$. 

This behavior is corroborated by the transfer -- gain or loss -- in the total (kinetic and potential) energy per unit time, shown in 
Fig.~\ref{fig:kinetic_e}(b). Here, as enlarged in the inset, we note the initial upward acceleration of particles due to the tap, followed by a 
coherent transfer of energy between layers from the bottom to the top, an effect similar to the ``stacked balls demonstration'' in introductory 
physics. The ballistic free-fall behavior leaves the total energy unchanged, which ends with a sequence of layer-by-layer crashes, 
progressively shorter and more intense, which dissipates the remaining kinetic energy we marked above as $E_{\rm eff}$. 

Figure~\ref{fig:kinetic_e}(c) shows $E_{\rm eff}$ as a function of $\Gamma$ for the same three layers represented in 
Fig. \ref{fig:kinetic_e}(a). Open symbols correspond to the continuous protocol following $\dot\Gamma_{3}$ while full symbols correspond 
to the $E_{\rm eff}$ obtained trough the stepwise protocol. It can be seen that $E_{\rm eff}$ is largely independent on the protocol 
followed. A black dashed-line represents $(A\omega)^{2}/(2gd)$, corresponding to the kinetic energy (in units of $mgd$) that a particle 
would acquire by moving with the maximum speed reached by the perturbation. The inset in the same figure shows a close-up of the main 
figure for lower values of $\Gamma$. The ``kink'' corresponds to the intensity at which particles in a given layer stop separating and the 
system moves as a solid. For example, for the bottom layers this happens at $\Gamma\approx1$, as expected for a solid without elastic 
interactions.
 
\paragraph{Data collapse as a function of $E_{\rm eff}$} 
In the following, we employ $E_{\rm eff}$ to collapse our data. To this end, we first consider the stepwise protocol. 
Figures \ref{fig:density} (a)-(b) show $\phi$ and $\Delta \phi$, respectively, for this protocol as a function of $\Gamma$, as in 
Fig.~\ref{fig:layers_gamma} but for all layers simultaneously. (Layers $2$, $3$, $8$, and $14$ are highlighted by color.)
From Fig. \ref{fig:density}(b) we notice that the sharply peaked form of $\Delta \phi$ as a function of $\Gamma$ is present in all the individual 
layers but occurs, however, at different $\Gamma$ values, as pointed out in Fig.~\ref{fig:layers_gamma}. Hence, when measured for the 
entire pile, these individual peaks in $\Delta \phi$ add up to form a ``trough'' as a function of $\Gamma$, as shown in the inset of 
Fig.~\ref{fig:density}(b). This behavior is consistent with previously reported results \cite{schroter2005stationary, ciamarra2006thermodynamics}.

Figures~\ref{fig:density}(c)-(d) show $\phi$ and $\Delta \phi$ as a function of $E_{\rm eff}$, respectively. Fig.~\ref{fig:density}(d) shows an 
almost perfect alignment of the peaks of $\Delta \phi$ as a function of $E_{\rm eff}$, supporting the hypothesis that $E_{\rm eff}$ is a 
physically relevant parameter to characterize the state of the system.  In particular, we mark $E_{\rm eff}=E_{\rm c}\approx 2$ as the 
effective energy where all fluctuations are simultaneously peaked as the ``transition energy". From Fig.~\ref{fig:density}(c), it can be seen 
that this new parameter ($E_{\rm eff}$) also aligns the inflection points for the densities $\phi$ in each layer in the same energy regime as 
the right slopes of $\Delta \phi$. 

Yet, a better collapse for the density in Fig.~\ref{fig:density}(c) is hindered by behavior resembling a ``finite-size'' effect for higher energies, 
where the data splays out systematically with the height of the layer. Although this residual effect is minor, we attempt to collapse this 
data further by rescaling with height $z$, defining the new scaling variable
\begin{equation}
 \tilde {E}=\left(E_{\rm eff} - E_{c} \right)/z
 \label{eq:scaling}
\end{equation}
This approach is entirely speculative and violates our desire to avoid system-specific parameters to unify the description. 
Fig.~\ref{fig:density}(e)-(f) show $\phi$ and $\Delta \phi$ as a function of $\tilde {E}$, respectively. While the collapse of the density $\phi$ in 
Fig.~\ref{fig:density}(e) indeed improves somewhat, the density fluctuations $\Delta \phi$ become progressively distorted in width for 
$z\to0$, see Fig.~\ref{fig:density}(f).
 
While this result appears to support the existence of a geometric constraint on the dynamics depending on the column height,  following the 
proposal in Refs.~\cite{gago2015relevance, pugnaloni2008nonmonotonic} this dependence could be explained using arguments based on the local expansion between particles 
mid-flight: From a configuration that would expand homogeneously, particles acquire a speed proportional to their relative height in the pile. 
Alternatively, this effect could also be associate with the contact network that particles form during flight, as has been discussed before 
\cite{gago2016ergodic}. There it is shown that the existence of persistent contacts (i.e., contacts that are never broken) during a perturbation is a source 
of memory in the system. These possible dependencies will be investigated in more detail in future work.

\begin{figure}
\centering \includegraphics[width=1\columnwidth]{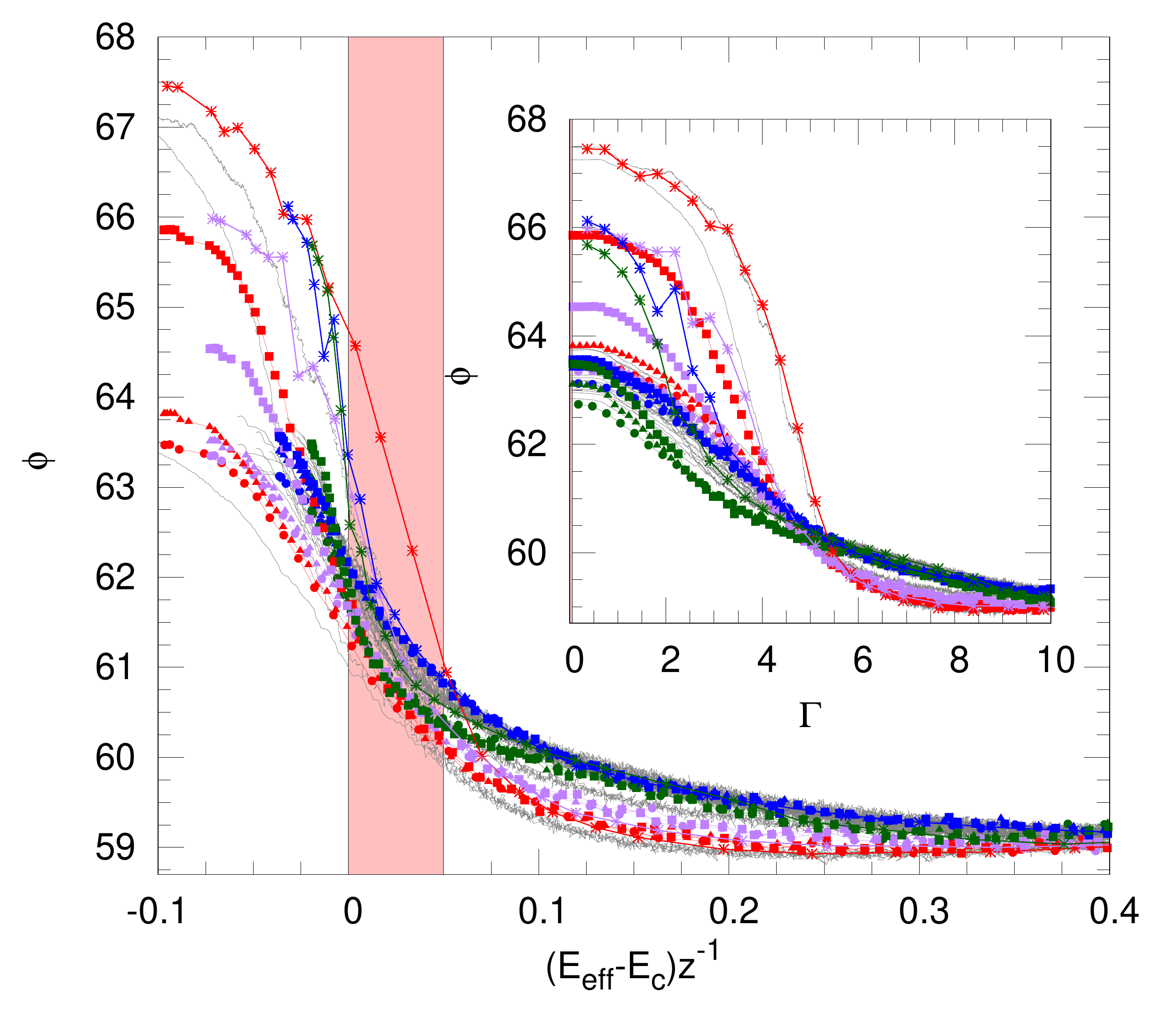}
\vskip -0.3cm
\caption{{\bf Data collapse for the continuous protocols.} (a) Density $\phi$ as a function of $ \tilde {E}$ (Eq. (\ref{eq:scaling})) as obtained for the stepwise (stars) and all three continuous annealing protocols: $\dot\Gamma_1$ (circles), $\dot\Gamma_2$ (triangles up) and $\dot\Gamma_3$ (squares), as well as for all layers (using different colours for the highlighted ones: Layer2 (red), Layer5 (purple), Layer8 (blue) and Layer14 (green)). Inset shows the same data but as a function of 
$\Gamma$. }
\label{fig:glass_transition} 
\end{figure}

Instead, we summarize all of our data for $\phi$ in Fig.~\ref{fig:glass_transition}. It demonstrates that there is a \emph{common} transition 
into a glassy state occuring in the region of energies corresponding to the right slope of the peak in $\Delta \phi$ vs $E_{eff}$ (highlighted in light-red). Above it, the scaling in Eq.~(\ref{eq:scaling}) collapses the data for the 
stepwise and all continuous protocols in all layers simultaneously, whereas for $E_{\rm eff}<E_{\rm c}$  the data is splitting into separate 
branches for different annealing rates $\dot\Gamma$ as well as between layers.
Thus, the critical value $E_{\rm c}$ provides a good 
prediction for the effective energy of particles at the glass transition. In contrast, the inset in Fig.~\ref{fig:glass_transition} illustrates the 
breadth over which these transitions spread out for $\Gamma$ as the controlling parameter, reinforcing the value of our new approach.

\paragraph{Conclusions}
In summary, we have shown that the density fluctuations, on approaching the glass transition from higher energies, rise to a sharp peak for 
each layer of the pile, before vanishing as the perturbation intensity decreases. By analyzing the grain-scale dynamics of the kinetic energy 
transfer and dissipation during the perturbation process, we have been able to define an effective kinetic energy that allows to quantify the 
energy of the transition in a unified manner for all layers and intensities. This is a groundbreaking statement, further justified by our ongoing 
studies on alternate settings discussed elsewhere, as prior experiments and simulations on granular piles were able to agree on qualitative 
results but struggle to achieve any quantitative comparison within the geometries and parameters they used. 

According to Fig.~\ref{fig:density}, the effective energy merely allows to align the data but does not achieve by itself a satisfactory collapse 
of $\phi$, hinting at other relevant quantities affecting the transition internally, which we are currently investigating. 
Although the ad-hoc re-scaling with the height $z$ reduces the apparent height-dependence and  produces a rather satisfactory collapse of the density data in Fig.~\ref{fig:density}(e) (see also Fig.~\ref{fig:glass_transition}), it unnaturally broadened the density fluctuations for $z\to0$ in Fig.~\ref{fig:density}(f). 
Nevertheless, it provides a direction for further investigation into the role of other grain-scale, dynamic parameters, such as the network of granular contacts, or the dependence of the dissipated energy on the mean-free-path (due to local expansion and frustration) among the grains.
To conclude, we remark that it has been shown that $\phi$ by itself is insufficient to completely characterize the (static, mechanically stable) configurations of a granular system \cite{pugnaloni2010towards, pugnaloni2011master, gago2015relevance, edwards2005full,blumenfeld2009granular,wanjura2020structural}. For this reason, further work will also have to address the behavior under the scaling provided by $E_{\rm eff}$ for other macroscopic quantities, such as the stress tensor of the system.

\section*{Materials and Methods}\label{Methods}
Figure~\ref{fig:setup} illustrates the setup for the granular pile used in our MD simulations. It consists of a cylindrical silo of 
diameter $D=2.4$cm with $60000$ spherical grains of slightly bi-dispersed diameters ($\left[1-1.02\right]$mm in equal number), to 
reduce crystallization. The height of the granular pile is $~12$cm, within a silo whose top (at $60$cm) was chosen high enough to ensure that the grains never interact with it. 
Within the $LIGGGHTS$ \cite{goniva2012influence} open source software implementation we also set a friction coefficient of $\mu=0.5$, a young modulus of $Y=10^{8}$ Pa, a restitution coefficient of 
$\epsilon=0.5$, a Poisson\textquoteright s ratio of $\nu=0.3$, and a density of $\rho=2500$kg m$^{-3}$ for our grains. The 
initial condition for the pile is obtained by simply pouring the grains into the container. 

The tap consists of a half sine-wave $A\sin(\omega t)$ with constant frequency 
$\omega=(2\pi / 0.047)$Hz, with the amplitude $A$ as the control parameter, applied to the silo by the movement of the entire container 
(both, bottom and side walls).
The dimensionless acceleration $\Gamma=A\omega^{2}/g$, with the gravity acceleration $g$, is used to represent the tap intensity.
Numerically, we consider that the system is static when the kinetic energy of the whole pile drops below a threshold of $10^{-1}$J.

\paragraph*{Measuring local density of static configurations:} 

To measure the packing fraction $\phi$, we divide the \emph{entire} system into $15$ cylindrical sub-regions ("layers"), as schematized in Fig.~\ref{fig:setup}(b), stacked along the height of the pile from $z=0.006875$ to $z=0.11$m.  To reduce boundary effects, particles closer than 
$2$mm from the silo lateral walls, $\approx 7$mm from the silo bottom, and particles on the surface ($z>0.11$m) are disregarded.
Each layer defined in this way contains $\approx2500$ particles.

Three different continuous rates of change were used: $\dot\Gamma_{1}=0.00002{\rm m}(\omega^{2}/g)$ per tap, $\dot\Gamma_{2}=\dot\Gamma_{1}/4$, and $\dot\Gamma_{3}=\dot\Gamma_{1}/8$. 
The packing fraction $\phi$ of each layer is measured after each tap. To this end, a Voronoi tesselation \cite{Rycroft2009} of the whole system is performed and the local density of 
each particle is obtained by dividing its volume by its corresponding Voronoi volume. The densities of those particles whose centers are in the 
sub-region of interest are averaged to obtain the packing fraction $\phi$ of the corresponding region. 

For the step-wise protocol we decrement the tap intensity by $\Delta A=0.00004$m for the first $7$ steps, and $\Delta A=0.00002$m for the remaining ones. 
For this rotocol, the selection of the number of taps applied at each given $A$ needed to satisfy the condition of the system reaching the stationary state at that $A$ as well as to provide enough statistics for an accurate calculation of the density fluctuations. Based on preliminary inspection of our data, we found that $500$ taps were sufficient.
To avoid the transient regime between consecutive intensities, the first $250$ taps at each $\Gamma$ are disregarded. Although this number is overly cautious at high tap intensities, it ensures that we only average over stationary states as the system evolves through its glassy phase.
For the remaining $250$ taps, we average over $\phi$, and $\Delta \phi$ is calculated as the standard deviation of the 
mean. 

\paragraph*{Measuring internal energies during the dynamics:} 
To measure the internal energy for each layer of the pile during the dynamics ensuing from the perturbation, we label particles with respect to 
the layers they reside in for the static configuration before the tap. We then track positions and velocities of all those particles during the dynamic process illustrated by Fig. \ref{fig:setup}(a) to calculate the average kinetic and gravitational potential energies per particle for each layer as a function of time.
From the decrease of mechanical energy, we can deduce the dissipation particles from the given layer have experienced during the process.

\acknowledgments These simulation were performed at the Imperial
College Research Computing Service (see DOI: 10.14469/hpc/2232).
Competing Interests: The authors declare that they have no competing interests.
All data needed to evaluate the conclusions in the paper are presented in the paper.
Author contributions: P.A.G. and S.B. designed research, performed research, analyzed data, and wrote the paper. 
P.A.G. and S.B. contributed equally to this work.
\medskip{}
\bibliographystyle{unsrt}
\bibliography{cites}

\end{document}